\documentclass[aps,prl,groupedaddress,notitlepage,twocolumn]{revtex4-1}
\usepackage{amsmath}
\usepackage{amsfonts}
\usepackage{graphicx}
\usepackage{color}
\usepackage{hyperref}

\newcommand{\ztwo}{\mathbb{Z}_2}
\newcommand{\ztwoC}{\mathcal{C}_{\ztwo}}

\begin{document}

\title{A bosonic topological phase in a paired superfluid}

\author{Snir Gazit, Ashvin Vishwanath}

\affiliation{Department of Physics, University of California, Berkeley, CA 94720, USA}

\date{\today}

\begin{abstract}
We study an effective model of two interacting species of bosons in two dimensions, which is amenable to sign problem free Monte Carlo simulations. In addition to conventional ground states, we access a paired superfluid which is also a topological phase, protected by the remaining $U(1) \times \ztwo$ symmetry. This phase arises from the condensation of a composite object, the bound state of vortices and anti-vortices of one species, to a boson of the second species. We introduce a bulk response function, the Ising analog of the quantized Hall effect, to diagnose the topological phase. The interplay of broken symmetry and topology leads to interesting effects such as  fractionally charged vortices in the paired superfluid. Possible extensions towards realistic models of cold atomic  bosons are discussed.

\end{abstract}


\maketitle

{\em Introduction -- }At low temperatures, interacting bosonic systems are expected to either be in a superfluid or an insulating phase. This classification follows from the conventional Landau theory which is based on symmetry breaking. Recently, it was realized that the insulating phase, in which the symmetry is preserved, can host additional phases characterized by non-trivial topological properties termed symmetry protected topological (SPT) phases \cite{ChenScience}. 

More broadly, SPT phases are defined through an equivalence relation between ground states of gapped Hamiltonians that are symmetric under a symmetry group $G$ and have no topological order (i.e., a unique ground state on a closed manifold). Two SPT phases are then said to be  in the same equivalence class if they can be adiabatically connected without breaking the symmetry. These considerations have led to a classification based on the Borel-group-cohomology \cite{SPT_PRB_Chen}, cobordism \cite{kapustin2014symmetry}, K-matrix theory  \cite{LuAshvinPRB} and non-linear sigma models \cite{NLSMZhen}.

Crucially, in the bosonic case, SPT phases are stable only in the presence of strong correlations since in the weakly interacting limit bosons inevitably condense. This is in sharp contrast to the non-interacting, band structure picture of electronic topological insulators.

Experimentally, the topologically protected spin one half edge states of the Haldane chain, a prime example of SPT phases in one spatial dimension \cite{PollmannSPT1D}, have been observed in neutron scattering experiments on Y$_2$BaNiO$_5$ compounds \cite{HaldanExp}. However, in spatial dimensions greater than one,  bosonic SPT phases have not been demonstrated experimentally yet.  In that regard, cold atomic systems, with their high flexibility in manipulating lattice structures and interactions \cite{BlochReview}, offer a promising experimental testbed for future realizations of SPT phases.

In certain cases, the physical mechanism underlying the SPT phase is based on real space binding of symmetry charges to topological defects \cite{ChenDW,XuSenthil}. A notable example is the bosonic quantum Hall (BQH) state  \cite{LuAshvinPRB,SenthiBQH} that can be realized by binding particles (holes) to vortex (anti-vortex) defects \cite{GeraedtsMotrunich}. Proliferating the charge decorated vortices gives rise to a topologically non-trivial insulating state characterized by a Hall conductance that is quantized to even integers $\sigma_{xy}=2\:n\: e^2/h$.  In a similar manner,  binding vortices to spin degrees of freedom yields a time reversal invariant SPT phase \cite{TRSPTWen}. 

Despite the appeal of the above mentioned approaches, current proposals pose experimental difficulties. More specifically, the BQH state breaks time reversal symmetry (TRS) and models realizing it \cite{RegnaultBQH,Ueda2013,he2015bosonic,SterdyniakOpticalFlux}  require a strong magnetic field or significant magnetic flux \cite{ArtGaugeFields} within a unit cell. In this extreme regime, the BQH state competes with fractional and even non-abelian topological phases. One may have hoped that realizing the BQH, a bosonic analog of the integer quantum Hall state, would be less demanding.

\begin{figure}[t]
	\includegraphics[scale=0.7]{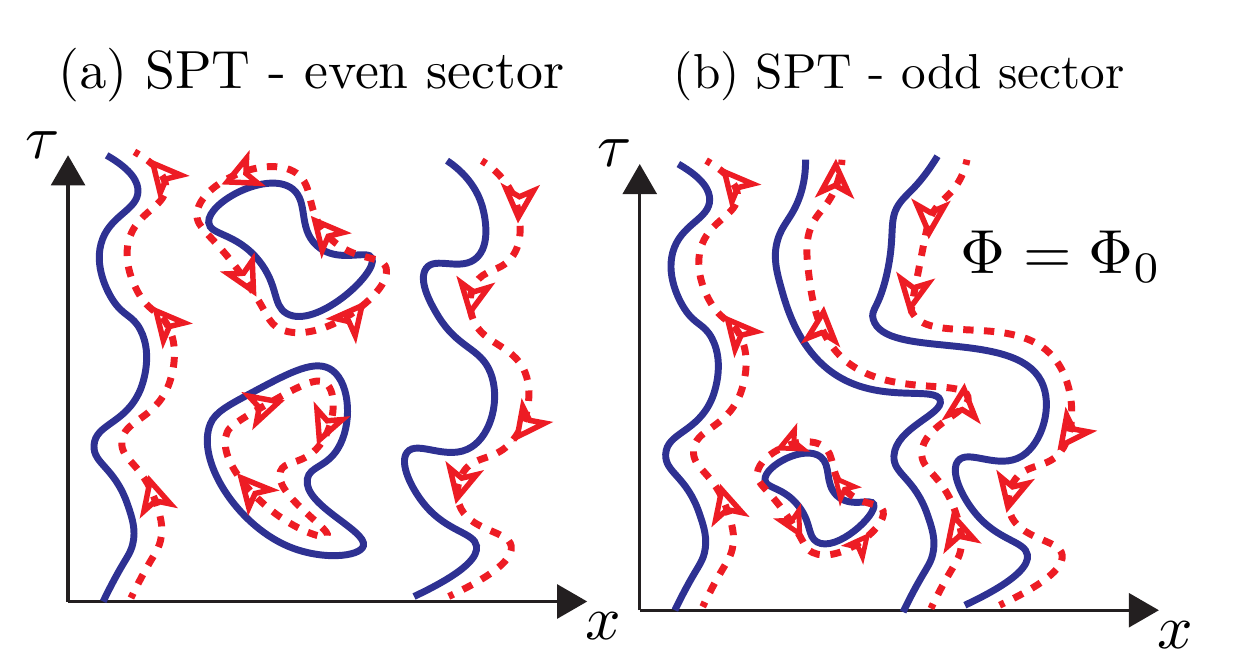}
	\caption{Typical world-line configurations in the SPT phase. Blue lines and red lines correspond to the $\ztwo$ world-lines and the vorticity current respectively. Threading a unit of flux quantum shifts the total $\ztwo$ charge  from the (a) even to the (b) odd sector.   }
	\label{fig:fig1}
\end{figure}

In this letter, we construct a two dimensional bosonic SPT phase that respects TRS and is composed solely of bosonic degrees of freedom. We implement our program of creating and identifying properties of this phase in an effective loop model using a sign problem free Monte Carlo (MC) simulation. To identify the SPT phase we introduce a procedure that directly measures the topological response of the SPT phase. In addition, we study the protected gapless edge states on a cylindrical geometry.

{\em Proposed construction --}  The construction is briefly summarized as follows. We begin with two species of bosons, labeled by A and B, with $U(1)\times U(1)$ symmetry corresponding to particle number conservation of each species separately. Let us assume that both species are in a superfluid phase. We then bind {\em both} the vortices and the anti-vortices of type A to the bosons of B, and condense this composite object. The choice of binding to bosons (rather than vacancies as in the BQH state) is crucial to ensure time reversal symmetry. 

Condensing the charged A vortices forms an insulator, while B bosons remain in a superfluid state. However, the superfluid is one of boson pairs, so the original symmetry is broken down to $U(1) \times \ztwo$. This residual symmetry protects a bosonic topological phase with nontrivial edge states.  

To see why this particular composite object condensate corresponds to a pair condensate, let us label the two composite objects that we are condensing as $\psi_+  = v_Ab_B$ and $\psi_-=v^\dagger_Ab_B$, where $v_A^\dagger$ ($v_A$) is the vortex creation (annihilation) operator and $b_B$ is the bosonic annihilation operator. Note, a condensate implies $\langle \psi_+\rangle,\,\langle\psi_-\rangle \neq0$. However, from this we cannot conclude that the B boson is condensed since vortices are nonlocal objects. The vorticity free combination that is condensed though is $\langle \psi_+\psi_-\rangle=\langle b_Bb_B\rangle\neq 0$, giving rise to a pair condensate.  

It is worth noting two points at this stage. First, the symmetric charge assignment may be argued to be easier to realize. The vortex core is associated with a reduced boson density.  Assuming a repulsive interaction between the two species of bosons, the vortex (and equally the anti-vortex) will be seen as a potential well for the opposite species of boson, potentially leading to a bound state. We caution that obtaining this binding is one of several ingredients required to create the desired phase. Second,  we note that the symmetry group protecting the SPT phase is a residual symmetry obtained by spontaneous symmetry breaking. This situation , to the best of our knowledge, has not been discussed before in the context of bosonic SPT phases. It provides a physical mechanism for introducing `gauge defects' that were suggested previously as theoretical devices to probe bosonic SPT phases  \cite{LevinGuGauge}. Here for example,  the pair condensate admits $\pi$ (or half) vortex defects, which are predicted to carry a half charge of the unbroken $U(1)$ symmetry of type A bosons.

Similarly to the trivial Bose insulator, condensing the $\ztwo$ charged vortices results in a $U(1)$ symmetric state \cite{PESKINDuality,DasguptaHalperin,FisherLeeCVD}. In addition, the $\ztwo$ symmetry is restored. This can be argued by noting that since the vortex is a non-local object one can not define a local order parameter for the bounded $\ztwo$ charges and thus the symmetry is preserved.

The above can also be understood by a  simple geometrical argument based on a world-line picture. Following the usual quantum to classical mapping\cite{polyakov1987gauge}, the partition function of a $\ztwo$ symmetric quantum system in $d$ spatial dimensions can be reformulated as a statistical mechanics model of unoriented loops (as opposed to $U(1)$ symmetry for which the loops are oriented) defined on a $d+1$ dimensional Euclidean space-time . In this language, the world-lines carry a $\ztwo$ charge and the total $\ztwo$ charge, $\ztwoC$,  equals to the parity of world-lines crossings at any given imaginary time slice. 

Let us recall the description of the conventional phases in this picture. In the disordered phase, the loop fugacity is small and hence a typical loop configuration consists out of small closed loops. In particular, winding around the imaginary time axis is suppressed by the finite single particle gap. As a result, the number of world-lines crossing, at any given imaginary time slice, is even and hence $\ztwoC=0$, as expected. By contrast, the ordered phase, where the loop fugacity is large, is characterized by large loops that can  wind around the imaginary time axis giving rise to fluctuations in $\ztwoC$ and breaking of the $\ztwo$ symmetry. 

Turning back to the non-trivial SPT phase, here the $\ztwo$ world-lines are bounded to the world-lines of the condensed vortices as depicted in Fig.~\ref{fig:fig1}a. Seemingly, the large $\ztwo$ loops could potentially lead to fluctuations of $\ztwoC$ and breaking of the $\ztwo$ symmetry. To understand why this is not the case, we note that on a closed manifold the total vorticity charge is neutral \cite{XYVortices}. This topological constraint restricts the number of vortex world-lines threading any imaginary time plane to be even. Consequently, also the total $\ztwo$ charge is even, i.e. $\ztwoC=0$, yielding a disordered state.

{\em SPT invariants -- }The symmetry is preserved both in the trivial and non-trivial SPT phases and hence they cannot be distinguished based on symmetry probes. The topological invariants characterizing different SPT phases are uncovered by gauging \cite{LevinGuGauge,MengTopInv,WenInvariant} the symmetry and examining the emergent topological field theory.  

To see how this applies in our case , it useful to first consider the minimal BHQ state with $\sigma_{XY}=2$, which at low energies can be described by an effective two component, $a_{I=A/B}$ ,mutual Chern-Simons (CS) theory\cite{LuAshvinPRB},
\begin{equation}
\mathcal{L}=\frac{1}{4\pi} \left(\epsilon^{\mu \lambda \nu} a_{A,\mu}\partial_\lambda a_{B,\nu}+A\leftrightarrow B\right).
\label{eq:mutual_Hall}
\end{equation}
Coupling the $U(1)$ currents, $J_I^\mu=\epsilon^{\mu \lambda \nu}\partial_\lambda a_{I,\nu}$,  to an external probe gauge fields $\mathcal{A}_{A/B,\nu}$  yields a quantized mutual Hall response, $J_{A/B}^\mu=\frac{1}{2\pi}\epsilon^{\mu \lambda \nu}\partial_\lambda \mathcal{A}_{B/A,\nu}$. Returning to our case, we now introduce a Higgs term \cite{MengTopInv} which breaks the $U(1)$ symmetry of type $B$ boson down to $\ztwo$, such that its charge is now identified only modulo two. The resulting topological response is then an Ising version of the quantum Hall effect, since threading an external unit of flux quantum, $\Phi_0=2\pi$, of the $U(1)$ symmetry generates a $\ztwo$ charge.

In terms of the  world-line picture, see Fig~\ref{fig:fig1}b, threading a unit of flux quantum induces one additional unit of vorticity that, in the SPT phase, carries a $\ztwo$ charge. As a result, the total $\ztwo$ charge shifts from the even to the odd sector, namely $\ztwoC=1$. Repeating the same analysis in the topologically trivial phase would have no effect since the $\ztwo$ charges are decoupled from the vortices.

An interesting consequence of the enlarged $U(1)\times U(1)$ symmetry is that the pair condensed state supports half-vortex excitation of type B bosons carrying one-half flux quantum $\Phi^B_{1/2}=\pi$ \cite{LeeHalfVortex}. Following the mutual Hall response in Eq.~\eqref{eq:mutual_Hall}, $j^0_A={\Phi^B_{1/2} / 2\pi}=1/2$, we see that a fractional one-half $U(1)$ charge of type B rotors is bounded to the $\pi$-vortex core.

{\em Coupled rotor model and observables -- }  To demonstrate numerically the above phenomenological approach,  we study a classical statistical mechanics model defined on a discrete 2+1 dimensional Euclidean space-time lattice. The degrees of freedom are two species of planar rotors parametrized by $\theta_A$ and $\theta_B$ that reside on the vertices, $r_i$  and $R_i$, of the direct and dual cubic lattice respectively. The partition function  is given by
\begin{equation}
\mathcal{Z}[\alpha_A,\alpha_B,\lambda]=\int \mathcal{D} \theta_{A} \mathcal{D} \theta_{B} f_{\alpha_A}(\theta_A) f_{\alpha_B}(\theta_B) g_{\lambda}(\theta_A,\theta_B).
\label{eq:part_func}
\end{equation}

\begin{figure}[!t]
	
	\hspace*{-0.5cm} 		\includegraphics[width=0.4\textwidth]{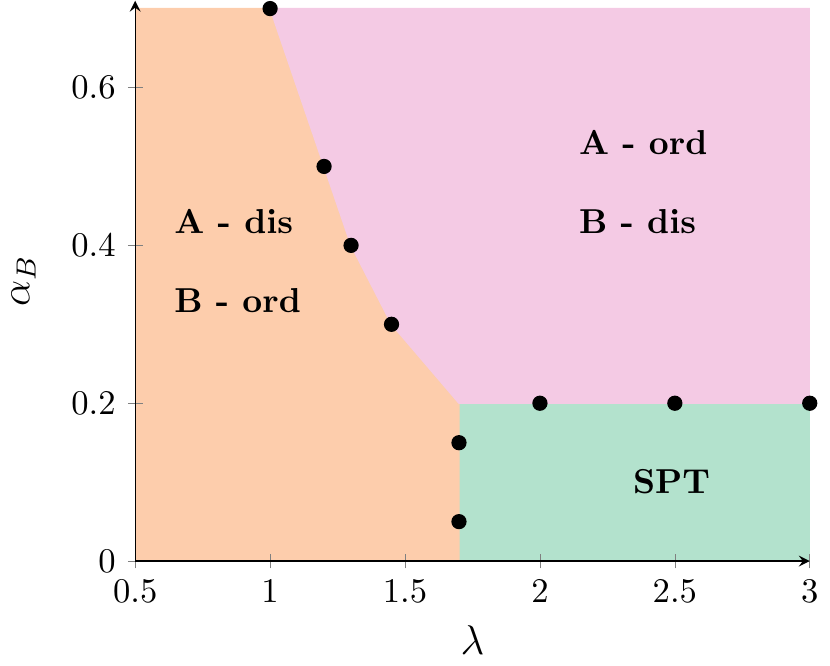}
	\caption{Phase diagram of the coupled rotor model in Eq.~\ref{eq:part_func} as a function of $\alpha_B$ and $\lambda$ for $\alpha_A=4$.  $U(1)$ symmetry breaking is probed by the condensate fraction and the phase transition points are marked by black circles. We label ordered (disordered) phases by "ord" ("dis"). }
	\label{fig:fig2}
\end{figure}
Here,  $f_\alpha(\theta)=\prod_{i,\mu} V_\alpha(\theta_i-\theta_{i+\hat{\mu}})$, with $\mu=x,y,\tau$, is a generalized XY model with nearest neighbor Boltzmann weight $V_\alpha(x)=1+2e^{-\alpha}\cos(x)$. The above three dimensional XY model captures the low energy properties of two dimensional lattice bosons with particle-hole symmetry (integer filling) \cite{FisherBosonLoc}. The atypical choice for the Boltzmann weight is designed such that in the dual loop current representation \cite{WallinDual} the integer bond currents $J_{i,\mu}$ are restricted to the values $J_{i,\mu}=0,\pm1$. In the quantum analogy, we allow only a single particle or hole excitation at each site.

The coupling between the rotor models, $g_\lambda(\theta_A,\theta_B)=e^{-\lambda \mathcal{S}_C[\theta_A,\theta_B]}$, is tuned by the coupling constant $\lambda$ and is defined through the binding action,
\begin{equation}
\mathcal{S}_C=\sum_{i} \left[ (|J^B_{R_i,\mu}|-|Q^A_{R_i,\mu}|)^2\right]
\end{equation}
Here, $J^B_{R_i,\mu}$ is the integer bond current of type B bosons, defined before, and  $Q^A_{R_i,\mu}$ is the vorticity three-current of type $A$ rotors. In defining the vorticity current we will consider the more general case where we thread a finite magnetic flux density $\phi=\Phi/L^2$, with $L$ being the linear system size. To do so, we minimally couple the bond current of type A rotors to an external gauge field $\mathcal{A}_{r_i,\mu}$ through a Peierls substitution, $\mathcal{J}_{r_i,\mu}^A=(\theta^A_{r_i}-\theta^A_{r_i+\hat{\mu}})\bmod 2\pi \to (\theta^A_{r_i}-\theta^A_{r_i+\hat{\mu}}-\mathcal{A}_{r_i,\mu})\bmod 2\pi$. The vorticity current is then given by the lattice curl, $2\pi Q^A_{R_i,\mu} +\phi_{R_i,\mu}= \nabla \times \mathcal{J}^A$. The magnetic flux density  $\phi_{R_i,\mu}=\phi\delta_{\mu,\tau}$ is uniform and it is non vanishing only along the imaginary time direction. The vorticity is quantized to integers and on a cubic lattice it is restricted to the values $Q^A_{R_i,\mu}=0,\pm1$

Possible symmetry breaking is probed through the condensate fraction, $C_{1/2}=\frac{1}{L^3}\sum_r g_{1/2}(r)$, where  $g_1(r)=\left\langle e^{i(\theta_i-\theta_{i+r})}\right\rangle $ and $g_2(r)=\left\langle e^{i2(\theta_i-\theta_{i+r})}\right\rangle $ are the single particle and pair correlation functions respectively. Pair condensation can also be detected from the winding number distribution, $P(W_\tau)$. The winding number, $W_\tau=1/L_\tau\sum_i J_{i,\tau}$, equals the total current along the imaginary time direction and can be interpreted as the total $U(1)$ charge in the quantum language. In a pair condensate, the probability for odd winding numbers vanishes, $P(W_\tau | W_\tau \text{ is odd} )=0$ \cite{PairSFWessel}.

A key ingredient of our analysis is identifying an order parameter that discriminates the non-trivial SPT phase from the trivial disordered phase. In our model the symmetry ($\ztwo)$ and its corresponding gauge field are discrete and hence linear response based observables, such as the quantized Hall conductance in the BQH case, cannot be defined. To resolve this, we follow the world-line picture by measuring $\ztwoC$ in the presence of a single flux quantum. A shift in $\ztwoC$ from the even to the odd sector serves as an order parameter that captures the topological response of this phase. In terms of the physical $U(1)$ degrees of freedom of type B rotors, $\ztwoC$  is defined through the winding number parity
\begin{equation}
\ztwoC=\left\langle\frac{1}{2}\left(1-(-1)^{W_\tau}\right)\right\rangle
\end{equation}

\begin{figure}[t]
	\hspace*{-0.5cm}                                                          
	\includegraphics[width=0.5\textwidth]{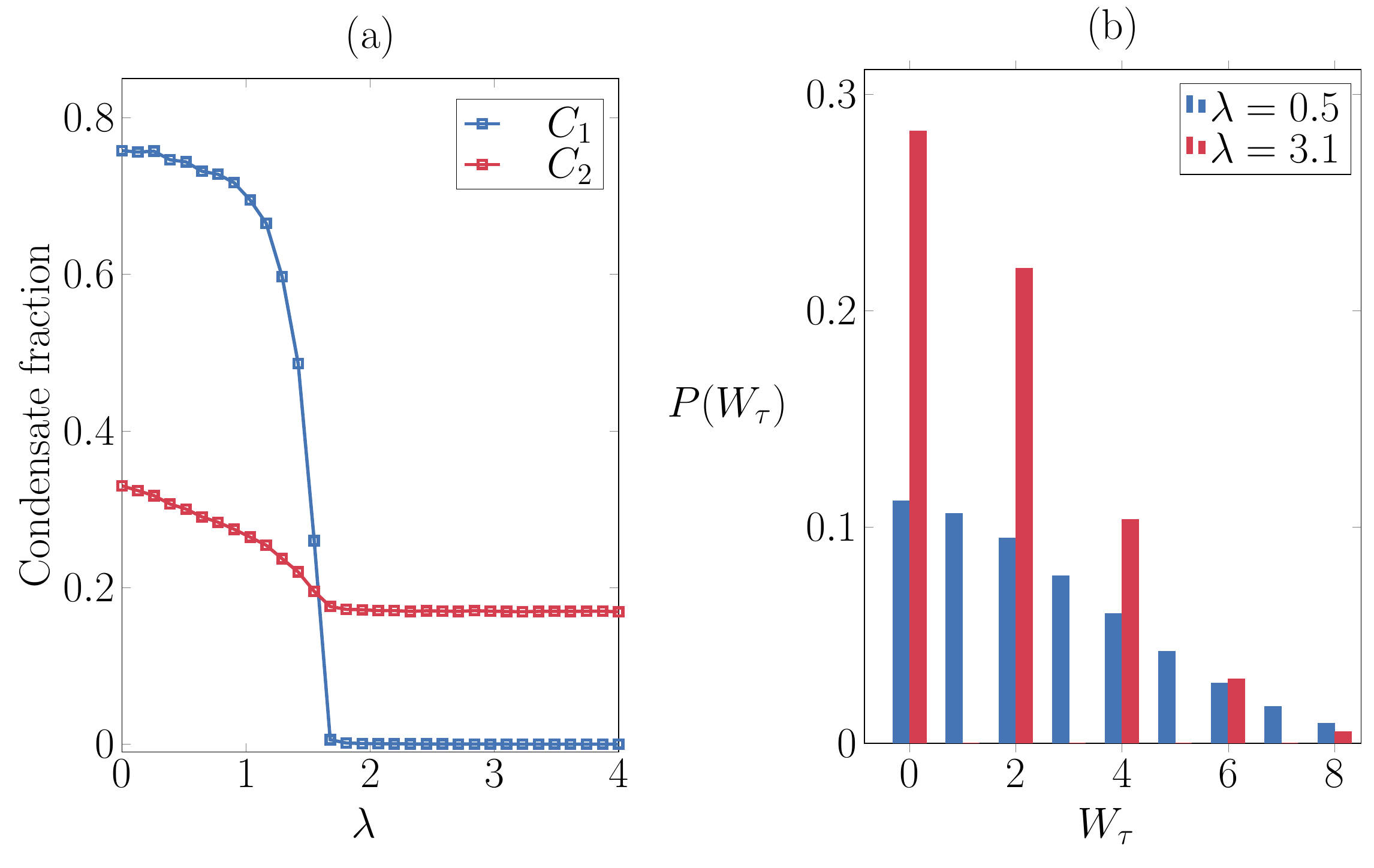}
	
	\caption{Signatures of pair condensation of type B rotors. For $\lambda>\lambda_c$, (a) the single particle condensate fraction, $C_1$, vanishes whereas the pair condensate fraction, $C_2$, remains finite and (b) the probability, $P(W_\tau)$, for odd winding numbers  vanishes.}
	\label{fig:fig3}
\end{figure}

{\em Methods -- } We evaluate the partition function in Eq.~\eqref{eq:part_func} by means of a classical MC. Type A (B) rotors were represented in the  phase (bond current) representation. The loop configuration of type B rotors is sampled using  the classical worm algorithm (WA) \cite{WorkProk}. We considered linear system size up to $L=32$. Further details on the MC algorithm can be found in the supplemental material \cite{supp}.

First, we determine the critical coupling of the decoupled partition function ($\lambda=0$),  to be $\alpha_c=1.4(1)$. We set  $\alpha_A=4>\alpha_c$ throughout such that in the decoupled limit type A rotors are disordered. This reduces the vortex fugacity and allows for the formation of the  $\ztwo$ charged vortices. 

{\em Results -- }  The numerically computed phase diagram as a function of $\alpha_B$ and  $\lambda$ is presented in Fig.~\ref{fig:fig2}. For weak binding (small  $\lambda$) type $B$ rotors undergo an ordering transition belonging the  usual three dimensional XY universality class as $\alpha_B$ is decreased. 

In the following we set $\alpha_B=0.05$ and study the phase diagram as a function of $\lambda$. In Fig.~\ref{fig:fig3}a  we depict the single particle and pair condensate fraction. At $\lambda_c=1.7(1)$, $C_1$ vanishes abruptly whereas $C_2$ remains finite signaling on the emergence of a pair condensate. 

We further verify this by studying the winding number distribution in Fig.~\ref{fig:fig3}b. For $\lambda=0.5<\lambda_c$, type B rotors are condensed as evident from the broad distribution of $P(W_\tau)$. For  $\lambda=3.1>\lambda_c$, while the distribution remains broad, the probably for odd winding numbers vanishes. This serves as a clear evidence for the formation of a pair condensate.  We also verified that type A rotors remain disordered in this parameter range.  Therefore, we conclude that the pair condensed phase has an unbroken $U(1) \times \ztwo $ symmetry and hence it is a candidate for a non-trivial SPT phase. 

\begin{figure}[t]
	\hspace*{-0.5cm}                                                           
	\includegraphics[width=0.5\textwidth]{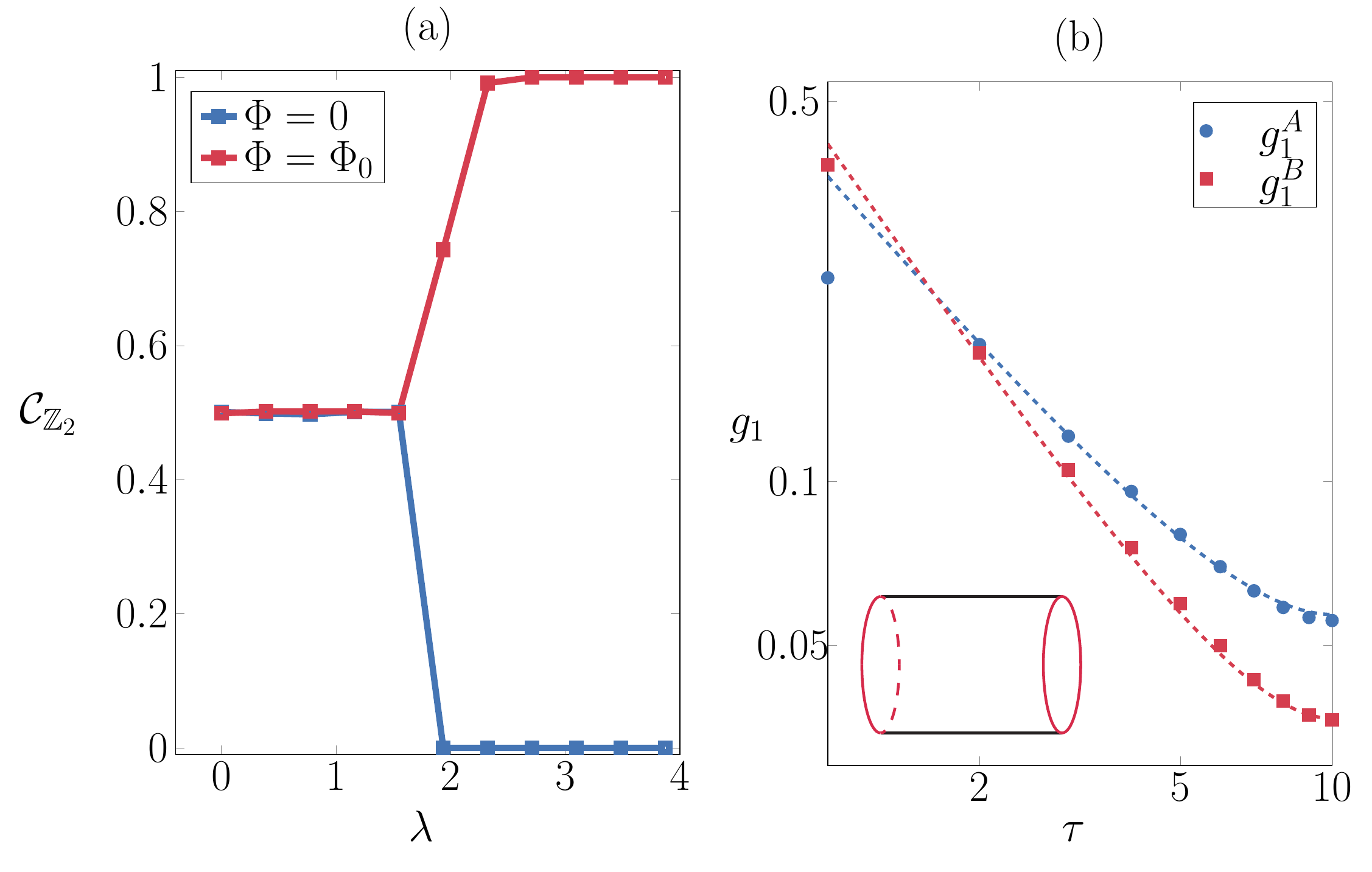}
	\caption{(a) The total $\ztwo$ charge, $\ztwoC$, as a function of $\lambda$ for $\Phi=0$ and $\Phi=\Phi_0$. In the SPT phase, $\ztwoC$ shifts from the even to the odd sector in response to an external unit of flux quantum. (b) Single particle Green's function, $g_1^{A/B}$, computed for $\lambda=2.5>\lambda_c$ along the edges of a cylinder ( $L_x=8,L_y=L_\tau=20$),  and presented in a log-log scale. Dashed lines correspond to a numerical fit to a power-law form.}
	\label{fig:fig4}
\end{figure}

Our main result is presented in Fig.~\ref{fig:fig4}a, where we depict $\ztwoC$ as a function of $\lambda$ both for $\Phi=0$ and $\Phi=\Phi_0$. For $\lambda<\lambda_c$,  type B rotors are condensed and thus the charge parity, $\ztwoC$, fluctuates and averages to one half. At $\lambda=\lambda_c$, $\ztwoC$ jumps abruptly to zero as the $\ztwo$ symmetry corresponding to the charge parity is restored. We now thread the torus with a single flux quantum, the charge parity rises to $\ztwoC=1$, precisely where  the pair condensate forms.  This provides a direct measurement of the topological response of the SPT phase.

Finally, we study the  edge states in the SPT phase on a cylindrical geometry . Gapless excitations are expected to follow an asymptotic power-law form proportional to $g_1^{A/B}(\tau)\sim  \left(\tau^{-\gamma_{A/B}}+(\beta-\tau)^{-\gamma_{A/B}}\right)$. We compute the single particle Green's function of both rotors along the edges of the cylinder for $\lambda=2.5$, as plotted in Fig.~\ref{fig:fig4}b. 

We numerically fit the MC data to the above form and  find good agreement with the exponents $\gamma_A=1.0(1)$ and $\gamma_B=1.3(1)$. Importantly, we have also explicitly verified that the bulk remains gapped. In the SPT phase, the two edge modes are conjugate variables \cite{GeraedtsMotrunich,LuAshvinPRB} and hence their Luttinger liquid parameters are related by a T-duality $\gamma_A \times \gamma_B=1$ \cite{fradkin2013field,supp}. Numerically we find, $\gamma_A \times \gamma_B\approx1.3(2)$. The small deviation from the analytic prediction is most likely related to finite size effects.

{\em Discussion and Summary -- }    Realizing a SPT phase in cold atomic systems requires better understanding of physical mechanisms leading to  binding of vortices to charge degrees of freedom, and condensation of  these composite objects. While we have used a loop model for convenience,in the future we anticipate microscopic implementations utilizing a lattice Hamiltonian. Translating terms from our loop model to the Hamiltonian formulation provides guidance along this direction.  For example, one promising approach is to introduce correlated hopping terms that were recently suggested in lattice realization of the BQH state \cite{he2015bosonic}.  We plan to apply this method to our model in a future study.
 
As a concrete experimental signature for the SPT phase, in cold atomic systems, we propose to probe the fractional one-half charges bounded to the $\pi$-vortices. Rotating the optical lattice can, in principle, induce vortices \cite{rotateVortex}, and the fractional charge can be measured by {\em in-situ} imaging \cite{InSituBloch}. We do not directly   demonstrate this effect in the numerical MC simulation since it introduces a sign problem.  

More generally, our numerical method for measuring the topological response of certain SPT phases protected by {\em discrete} symmetry in MC simulations can be applied to other examples of SPT phases with different symmetries and dimensionality. 

Summarizing, we proposed a purely bosonic model that following pair condensation realizes a SPT phase protected by $U(1) \times \ztwo$ symmetry. The signatures of the SPT phase were probed in an effective lattice model, and the interplay with spontaneous symmetry breaking was discussed.  Our approach could guide the search for possible realizations of SPT phases in realistic models.

{\em Acknowledgments -- }
We thank Itamar Kimchi, Ehud Altman, Mike Zaletel, Yuan Ming Lu and Norman Yao for discussions, and acknowledge support from the Templeton Foundation and AFOSR MURI grant
FA9550-14-1-0035.

\bibliography{spt_z2_u1}{}

\newpage
\onecolumngrid

\begin{center}
	\textbf{\large Supplemental Material: ``A bosonic topological phase in a paired superfluid"}
\end{center}
\setcounter{equation}{0}
\setcounter{figure}{0}
\setcounter{table}{0}
\makeatletter
\renewcommand{\theequation}{S\arabic{equation}}
\renewcommand{\thefigure}{S\arabic{figure}}
\renewcommand{\bibnumfmt}[1]{[S#1]}
\renewcommand{\citenumfont}[1]{S#1}

\section{Single particle Green's function for  general edge interactions}

The low energy description of the edge states in the SPT phase is given by the action
\begin{equation}
\begin{aligned}
\mathcal{L}&=\frac{1}{4\pi} \int dx dt \left( \partial_t\phi_1\partial_x\phi_2
+\partial_t\phi_2\partial_x\phi_1
+ \sum_{I,J=1,2} V_{I,J} \partial_x\phi_I\partial_x\phi_J\right)
\end{aligned}
\end{equation}

The matrix $V_{I,J}$ is not universal and corresponds to interactions between the edge modes. The $K=\begin{pmatrix}
0&1\\1&0
\end{pmatrix}$ and  $V_{IJ}$ matrices can be diagonalized simultaneously since $K$ is symmetric and $V$ is symmetric and positive definite  \cite{FisherLLSPT}. Explicitly, we define
\begin{equation}
\begin{aligned}
\phi_1&=\frac{1}{\sqrt{2g}}\left(X_L+X_R\right) \\ \nonumber
\phi_2&=\sqrt{g/2}\left(X_R-X_L\right) 
\end{aligned}
\end{equation}
\begin{equation}
\begin{aligned}
\mathcal{L}& =\frac{1}{4\pi} \int dx dt\partial_x X_R\left( \left(\partial_t-v_R\partial_x\right)X_R + \partial_x X_L\left(-\partial_t-v_L\partial_x\right) X_L\right)
\end{aligned}
\end{equation}
The single particle Green's function is then given by
\begin{equation}
\begin{aligned}
\left\langle e^{i(\phi_1(x,t)- \phi_1(0,0))}\right\rangle&\propto \prod_{R/L} |x-v_{R/L} t|^{g}\\ \nonumber
\left\langle e^{i(\phi_2(x,t)- \phi_2(0,0))}\right\rangle&\propto \prod_{R/L}|x-v_{R/L}t|^{1/g}
\end{aligned}
\end{equation}

From the above equations we see that the product of the Luttinger liquid parameters is unity. 

\section{Implementation details of the Monte Carlo algorithm }

In this section we provide a detailed description of the MC algorithm used to compute the classical partition function in Eq.~\ref{eq:part_func}. As mentioned in the main text, we reformulated the action of type B rotors as an integer bond current model. The closed loop configurations is sampled using the worm algorithm \cite{WorkProk,WolffReweight}. In the pair condensed phase, worm updates of a single field insertion, $e^{i\theta}$, are inefficient due to the finite single particle gap. To address this we introduced updates in which the worm's head  carries two field insertions, $e^{i2\theta}$. This also enabled us to directly measure the pair correlation function \cite{PairSFWessel}. Type B rotors were represented by an angle variable and we employed the usual Metropolis-Hastings single site update  scheme.

In the SPT phase, where the coupling $\lambda$ is sizable, MC moves that sample the A and B rotors separately are inefficient since they must overcome an energy barrier in order to generate a bound state of a $\ztwo$ charge and a vortex. To overcome this difficulty we introduced the following MC move. First, we propose an angle update to a type A rotor at a randomly selected site $r_i$ ( belonging to the direct lattice). Such a move can potentially generate a vortex loop. We then randomly select a direct lattice bond $b(r_i)$ out of the bonds emanating from the site $r_i$. Finally, we suggest to construct a loop current of type B rotors surrounding the bond $b(r_i)$ in the dual lattice. The move is accepted or rejected according the {\em total} Boltzmann ratio. We found that this simple move allows the formation of $\ztwo$ charged vortex loops and significantly reduces the MC correlation time.

\end{document}